\documentclass[prl,reprint,twocolumn,showpacs,superscriptaddress,floatfix,aps,10pt]{revtex4-2}

\usepackage{amsmath,amsthm,amssymb}
\usepackage{dcolumn,bm,hyperref}
\usepackage{graphicx,subfigure,verbatim}
\usepackage{newtxtext}
\usepackage{xcolor}
\usepackage{soul}
\usepackage{ulem}

\usepackage{xcolor}

\begin{document}

\title{Magnetic Octupole Hall Effect in $d$-Wave Altermagents
}
\author{Hye-Won Ko}
\affiliation{Department of Physics, Korea Advanced Institute of Science and Technology (KAIST), Daejeon 34141, Korea}
\author{Kyung-Jin Lee}
\email{kjlee@kaist.ac.kr}
\affiliation{Department of Physics, Korea Advanced Institute of Science and Technology (KAIST), Daejeon 34141, Korea}

\begin{abstract}
Order parameters not only characterize symmetry-broken equilibrium phases but also govern transport phenomena in the nonequilibrium regime. Altermagnets, a class of magnetic systems integrating ferromagnetic and antiferromagnetic features, host multipolar orders in addition to dipolar N{\'e}el order. In this work, we demonstrate the multipole Hall effect in $d$-wave altermagnets---a transverse flow of multipole moments induced by an electric field. Using symmetry analysis and linear response theory, we show that the magnetic octupole Hall effect persists even in symmetries where the spin-splitter effect is forbidden and thus provides a robust experimental signature. In addition, we identify a sizable electric quadrupole Hall effect, originating from quadrupole splittings in the band structure. Our results expand the family of Hall effects to include higher-order multipolar responses and establish altermagnets as a versatile platform for exploring multipole transport beyond spin and orbital degrees of freedom.
\end{abstract}

\maketitle

{\it Introduction.}---Antiferromagnetism has traditionally been understood as the antiparallel alignment of magnetic dipole moments~\cite{Neel1936}, in contrast to the parallel alignment in ferromagnets.
Recently, altermagnets~\cite{Smejkal2022} have garnered significant attention for exhibiting broken time-reversal symmetry despite having zero net magnetization~\cite{Smejkal2020,Reichlova2024}. This seemingly incompatible coexistence of ferromagnetic and antiferromagnetic charateristics is reconciled by the presence of ferroic magnetic multipolar order~\cite{Bhowal2024,McClarty2024,Verbeek2024}, in addition to the antiferroic N{\'e}el order. In particular, the ferroic multipolar order breaks the time-reveral symmetry and dictates the symmetry of spin-split electronic bands~\cite{Yuan2020,Lee2024,Krempasky2024,Reimers2024}, a key feature of altermagnets. Multipolar degrees of freedom are thus essential for fully describing the nature and potential of altermagnetism.

Spontaneous multipole orderings have been extensively studied in strongly correlated $f$-electron systems~\cite{Kusunose2008,Kuramoto2009,Santini2009} as a route to exotic phases beyond conventional ferromagnetism and ferroelectricity. Aspherical distributions of charge and magnetization densities—arising from the interplay of Coulomb interactions, spin-orbit coupling (SOC), and crystal field effects—are characterized by nonvanishing electric and magnetic multipole moments, respectively. 
The scope of multipolar phases has expanded to include $d$-orbital systems with strong SOC~\cite{Chen2010,Fu2015,Harter2017,Hirai2020}.
In these systems, multipolar orders and their fluctuations can give rise to various emergent phenomena, including unconventional superconductivity~\cite{Sakai2012,Matsubayashi2012,Tsujimoto2014,Sumita2017} and multipolar Kondo effect~\cite{Cox1988,Patri2020}, further enriching the landscape of multipole physics.

Identifying order parameters not only classifies equilibrium phases but also provides a natural framework for predicting cross-correlated responses~\cite{Hayami2018a,Yatsushiro2021,Watanabe2018}. Once the order parameter is specified, nonvanishing components of response tensors can be systematically identified by decomposing them into symmetry-adapted multipoles~\cite{Kusunose2023}. In particular, multiferroic responses, such as magnetoelectric or piezoelectric effects, are anticipated when the spontaneous multipolar order shares the symmetry of the corresponding response tensors. For instance, isotropic volume change can be induced by an electric field in the presence of an electric dipole, since the relevant piezoelectric tensor is a time-reversal-even, rank-1 polar tensor—exhibiting the same symmetry as the electric dipole~\cite{Hayami2018a}.
Despite this established framework for nonequilibrium responses based on multipoles, the transport of multipole moments remains largely unexplored~\cite{Tahir2023}. In conventional ferromagnets, assuming negligible SOC, magnetic dipole order defines spin as a good quantum number for each $\bf k$-state, leading to spin-polarized currents under an applied electric field—forming the basis of spintronics. Analogously, in compensated antiferromagnets with N\'eel order, electric fields can generate staggered N\'eel spin currents~\cite{Shao2023}. These direct connections between order parameters and current responses suggest that multipolar currents should naturally arise in systems with multipole order when subjected to an electric field.

In this Letter, we demonstrate the multipole Hall effect in nonrelativistic $d$-wave altermagnets, characterized by magnetic octupole order. Based on symmetry analysis and linear response theory, we show that the magnetic octupole Hall effect emerges regardless of whether the spin-splitter effect~\cite{Hernandez2021} is symmetry-allowed or forbidden. The persistence of the magnetic octupole Hall effect even in the absence of the spin-splitter effect highlights that magnetic octupole responses are inherent to transport in $d$-wave altermagnets.
This finding extends the conventional perspective on altermagnetic phenomena, which have so far focused primarily on spin degrees of freedom, and highlights the essential role of multipolar order in governing both equilibrium properties and nonequilibrium responses in altermagnets.

\begin{figure}[t]
\begin{center}
\includegraphics[scale=1.0]{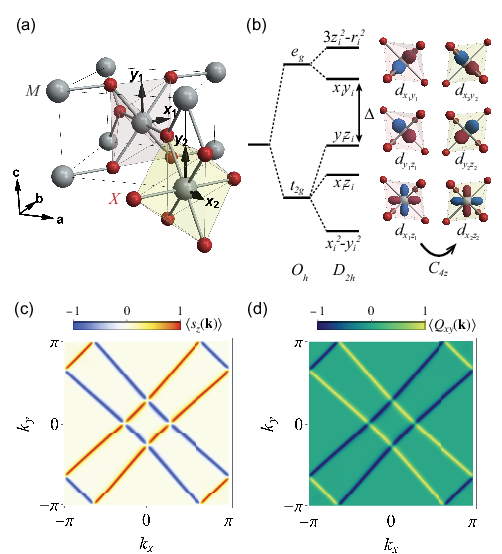}
\caption{
(a) Crystal structure of tetragonal rutile $MX_2$, where gray and red spheres represent the magnetic component $M$ and the nonmagnetic component $X$, respectively. Two magnetic sublattices $M_i$ $(i=1,2)$ are located at the volumetric center and corner, with their local coordinate axes denoted as $({\bf x}_i,{\bf y}_i,{\bf z}_i)$.
(b) Energy-level diagram of local $d$ orbitals for $D_{2h}$ symmetry~\cite{Occhialini2021}. The right figures illustrate the local $d$ orbitals at the magnetic sublattice $M_i$---$d_{x_iy_i}$, $d_{y_iz_i}$, and $d_{x_iz_i}$---viewed along the $[001]$ direction. The upper two orbitals, $d_{x_iy_i}$ and $d_{y_iz_i}$, are split by $\Delta$. 
(c,d) Normalized Fermi surface expectation values of (c) the spin-$z$ component $\langle s_z({\bf k}) \rangle$ and (d) the electric quadrupole $\langle Q_{xy}({\bf k}) \rangle$, calculated from the Hamiltonian in Eq.~(\ref{Hbulk}) at Fermi energy $E_{\rm F}\!=\!-3.0~{\rm eV}$. 
}
\label{fig:1}
\end{center}
\end{figure}

{\it Model Hamiltonian.}—We consider rutile compounds $MX_2$ [Fig.~\ref{fig:1}(a)], proposed as candidates for $d$-wave altermagnets~\cite{Bhowal2024,Smejkal2020}. In these systems, the two magnetic sublattices are related by a fourfold rotation $C_{4z}$ [Fig.~\ref{fig:1}(b)], combined with time-reversal symmetry. This symmetry originates from distorted $MX_6$ octahedra, which lower the local crystallographic symmetry at the transition metal $M$ sites from $O_h$ to $D_{2h}$ in a sublattice-dependent manner~\cite{Occhialini2021}, thereby lifting the degeneracy of the $t_{2g}$ and $e_g$ orbitals [see crystal field splitting in Fig.~\ref{fig:1}(b)]. 
Partial occupation of these nondegenerate orbitals imparts anisotropic orbital character to the $d$-electrons, leading to aspherical charge distributions around the $M$ ions. In rutile-type $\rm MnF_2$~\cite{Bhowal2024}, for example, the staggered structural distortion gives rise to an antiferroic electric quadrupole order, reflecting the local $d_{y_iz_i}$ orbital character [Fig.~\ref{fig:1}(b)], while the antiferromagnetic spin order follows the same sublattice pattern. The combination of these two antiferroic orders induces a ferroic magnetic octupole order, defined as the product of spin and electric quadrupole moments, even in the absence of SOC. The onset of altermagnetism, characterized by the ferroic magnetic multipole order, thus stems from the concurrent emergence of spin and orbital orderings, where the latter can be stabilized either by crystal symmetry or electronic instability~\cite{Leeb2024}.

We construct a tight-binding model for altermagnetic rutile $MX_2$ that incorporates the sublattice-dependent crystal field effects described above.
To accommodate various transition-metal $M$ ions—such as $\rm Ru^{4+}$ in $\rm RuO_2$ and $\rm Mn^{2+}$ in $\rm MnF_2$—we adopt the $d_{yz}$, $d_{xz}$, and $d_{x^2\!-\!y^2}$ orbitals, defined in the global Cartesian coordinate system $\bf (x,y,z)\!=\!(a,b,c)$, as a minimal basis set. These orbitals transform into the local $d_{x_iz_i}$, $d_{y_iz_i}$, and $d_{x_iy_i}$ orbitals on each $M_i$ $(i=1,2)$ sublattice via a unitary transformation [see Fig.~\ref{fig:1}(b)]. Compared to previous models~\cite{Bhowal2024,Vila2025}, which account for local crystal field effects within two-dimensional orbital space, the chosen set of orbitals not only captures the lowered crystallographic symmetry but also forms a minimal yet complete basis that fully encapsulates electric quadrupoles, and concomitant magnetic octupoles in magnetic phase. 

The full Hamiltonian is constructed in the Hilbert space spanned by the product of spin, sublattice, and orbital bases, namely: $\{|\!\uparrow\rangle,|\!\downarrow\rangle\}\otimes\{|M_1\rangle,|M_2\rangle\}\otimes\{|d_{yz}\rangle,|d_{xz}\rangle,|d_{x^2\!-\!y^2}\rangle\}$. In this basis, the Hamiltonian takes the form:
\begin{align}
    \mathcal{H} = \sigma_0\otimes\mathcal{H}_0+J\sigma_z\otimes\tau_z\otimes\mathbb{I}, 
    \label{Hbulk}
\end{align}
where $\sigma_\mu$ and $\tau_\mu$ $(\mu\!=\!0,x,y,z)$ are Pauli matrices acting in spin and sublattice spaces, respectively, $\mathbb{I}$ is the identity matrix in orbital space, and $J$ denotes the exchange splitting. SOC is excluded to isolate the nonrelativistic origin of altermagnetic phenomena. 
The spin-independent Hamiltonian $\mathcal{H}_0$ is given by (see Supplemental Material~\cite{SM} for further details)
\begin{align}
    \mathcal{H}_0 = \sum_{\mu=0,x,z}\sum_{i,j=x,y,z} h^{ij}_\mu({\bf k})~\tau_\mu\otimes\{L_i,L_j\}, \label{H0}
\end{align}
where $h^{ij}_\mu({\bf k})$ are $\bf k$-even functions and $L_i$ denotes the orbital angular momentum operator. Here, the symmetrized product of orbital angular momentum operators $\{L_i,L_j\}$, referred to as orbital angular position~\cite{Han2022}, is equivalent to electric quadrupoles (see below).
The explicit form of $h^{ij}_\mu({\bf k})$ is listed in Supplemental Material~\cite{SM}, considering up to the next-nearest neighbor hopping. We set $h^{xy}_z({\bf k})$ as the orbital splitting $\Delta$, i.e., $\Delta\tau_z\otimes\{L_x,L_y\},$ which encodes the sublattice-dependent crystal field effects arising from the reduced local symmetry. This term lifts the degeneracy between $d_\pm(\equiv~d_{yz}\pm d_{xz})$ orbitals---corresponding to the local $d_{x_iy_i}$ and $d_{y_iz_i}$ orbitals shown in Fig.~\ref{fig:1}(b).

\begin{table*}[t]
\centering
\caption{Symmetry-allowed components of the ${\cal O}$-conductivity tensors $\sigma^{\cal O}_{ij}$ [Eq.~(\ref{Kubo})] for the spin point group $^24/^1m^2m^1m$, where ${\cal O}$ denotes spin ($s_i$), electric quadrupole ($Q_{ij}$), or magnetic octupole ($M_{ijk}\!=\!s_iQ_{jk}$). The first row indicates the symmetry-restricted form of electric quadurpole conductivity tensors, while the second row lists the spin and magnetic octupole conductivity tensors. Cubic harmonic labels are used to identify the five irreducible rank-2 electric quadrupoles, and also serve as shorthand for the spatial indices of the magnetic octupoles. 
The Cartesian coordinate system corresponds to Fig.~\ref{fig:1}(a), with $\bf x\!=\!a$, $\bf y\!=\!b$, and $\bf z\!=\!c$.}
\label{TensorSpin}
\setlength{\tabcolsep}{1.5pt} 
\renewcommand{\arraystretch}{1.3} 
\begin{tabular}{ccccccc}
\noalign{\smallskip}\noalign{\smallskip}
\hline\hline
 & $\sigma^{Q_u}$ & $\sigma^{Q_v}$ & $\sigma^{Q_{yz}}$ & $\sigma^{Q_{xz}}$ & $\sigma^{Q_{xy}}$ \\
\hline
 &
$\begin{pmatrix} \sigma^{Q_u}_{xx} & 0 & 0 \\ 0 & \sigma^{Q_u}_{xx} & 0 \\ 0 & 0 & \sigma^{Q_u}_{zz} \end{pmatrix}$ &
$\begin{pmatrix} \sigma^{Q_v}_{xx} & 0 & 0 \\ 0 & -\sigma^{Q_v}_{xx} & 0 \\ 0 & 0 & 0 \end{pmatrix}$ &
$\begin{pmatrix} 0 & 0 & 0 \\ 0 & 0 & \sigma^{Q_{xz}}_{xz} \\ 0 & \sigma^{Q_{yz}}_{zy} & 0 \end{pmatrix}$ &
$\begin{pmatrix} 0 & 0 & \sigma^{Q_{xz}}_{xz} \\ 0 & 0 & 0 \\ \sigma^{Q_{yz}}_{zy} & 0 & 0 \end{pmatrix}$ &
$\begin{pmatrix} 0 & \sigma^{Q_{xy}}_{xy} & 0 \\ \sigma^{Q_{xy}}_{xy} & 0 & 0 \\ 0 & 0 & 0 \end{pmatrix}$ \\
\hline
$\sigma^{s_z}$ & $\sigma^{M_{zu}}$ & $\sigma^{M_{zv}}$ & $\sigma^{M_{zyz}}$ & $\sigma^{M_{zxz}}$ & $\sigma^{M_{zxy}}$ \\
\hline
$\begin{pmatrix} 0 & \sigma^{s_z}_{xy} & 0 \\ \sigma^{s_z}_{xy} & 0 & 0 \\ 0 & 0 & 0 \end{pmatrix}$ &
$\begin{pmatrix} 0 & \sigma^{M_{zu}}_{xy} & 0 \\ \sigma^{M_{zu}}_{xy} & 0 & 0 \\ 0 & 0 & 0 \end{pmatrix}$ &
$\begin{pmatrix} 0 & \sigma^{M_{zv}}_{xy} & 0 \\ -\sigma^{M_{zv}}_{xy} & 0 & 0 \\ 0 & 0 & 0 \end{pmatrix}$ &
$\begin{pmatrix} 0 & 0 & \sigma^{M_{zyz}}_{xz} \\ 0 & 0 & 0 \\ \sigma^{M_{zxz}}_{zy} & 0 & 0 \end{pmatrix}$ &
$\begin{pmatrix} 0 & 0 & 0 \\ 0 & 0 & \sigma^{M_{zyz}}_{xz} \\ 0 & \sigma^{M_{zxz}}_{zy} & 0 \end{pmatrix}$ &
$\begin{pmatrix} \sigma^{M_{zxy}}_{xx} & 0 & 0 \\ 0 & \sigma^{M_{zxy}}_{xx} & 0 \\ 0 & 0 & \sigma^{M_{zxy}}_{zz} \end{pmatrix}$ \\
\hline\hline
\end{tabular}
\end{table*}

To characterize the resulting multipole orders, we introduce electric and magnetic multipole operators. The rank-$l$ electric multipole operator $Q^m_l$ is defined as $Q^m_l\!=\!-e\sum_j \sqrt{4\pi/(2l+1)}r^lY^m_l({\bf r}_j)$ $(-l\!\le\! m\!\le\! l)$~\cite{Kusunose2008,Hayami2018a,Hayami2018b,Kusunose2020}, where $Y^m_l({\bf r})$ are spherical harmonics and the sum runs over all electrons at positions ${\bf r}_j$. For quadrupoles ($l\!=\!2$), this expression can be recast using cubic harmonics as $Q_u\!\equiv\!Q_{3z^2-r^2}\!\equiv\!Q^0_2$, $Q_v\!\equiv\!Q_{x^2-y^2}\!\equiv\!(Q^{-2}_2\!+\!Q^2_2)/\sqrt{2}$, $Q_{yz}\!\equiv\!i(Q^{-1}_2\!+\!Q^1_2)/\sqrt{2}$, $Q_{xz}\!\equiv\!(Q^{-1}_2\!-\!Q^1_2)/\sqrt{2}$, and $Q_{xy}\!\equiv\!i(Q^{-2}_2\!-\!Q^2_2)/\sqrt{2}$. According to the Wigner-Eckart theorem~\cite{Inui1990}, matrix representations of electric quadrupole operators can be constructed from symmetrized products of angular momentum operators as $Q_{ij}\propto\{L_i,L_j\}-\delta_{ij}/3$.
The correspondence between electric quadrupole and orbital angular position operators suggests that the antiferroic $Q_{xy}$ ordering~\cite{Bhowal2024} is integrated by the orbital splitting $\Delta$.
Magnetic octupole moments, defined in Cartesian indices as $M_{ijk}\!=\!\int d{\bf r}~m_i({\bf r})r_jr_k$, describe the spatial distribution of the magnetization density $\bf m(r)$~\cite{Urru2022}. Assuming that the magnetic moment arises solely from spin, the magnetic octupole operator is defined as
\begin{equation}
    M_{ijk}=s_iQ_{jk},
\end{equation}
where $s_i$ ($i\!=\!x,y,z$) are the dimensionless spin operator. Rather than employing irreducible rank-3 representations, we adopt this reducible form to explicitly separate spin and spatial sectors~\cite{SM}.

Figures~\ref{fig:1}(c) and \ref{fig:1}(d) respectively show the spin and electric quadrupole expectation values, $\langle s_z({\bf k})\rangle$ and $\langle Q_{xy}({\bf k})\rangle$, at the Fermi surface for $E_{\rm F}\!=\!-3.0~{\rm eV}$, computed from the Hamiltonian in Eq.~(\ref{Hbulk}).
Two key features are evident. First, the electronic bands exhibit not only spin splitting [Fig.~\ref{fig:1}(c)] but also electric quadrupole splitting [Fig.~\ref{fig:1}(d)]. Second, the spin- and electric quadrupole-split bands are exactly overlapped in momentum space, with $\langle s_z({\bf k})\rangle$ and $\langle Q_{xy}({\bf k})\rangle$ exhibiting identical $\bf k$-dependent sign changes. Consequently, each $\bf k$-state carries a definite spin and a definite electric quadrupole moment, and their product, $\langle s_z({\bf k})Q_{xy}({\bf k})\rangle$, realizes a ferroic magnetic octupole order $M_{zxy}$~\cite{Bhowal2024}. This one-to-one correspondence originates from the spatially aligned spin and orbital orderings mediated by the sublattice degrees of freedom.

{\it Symmetry analysis of multipole conductivity tensors.}—The antiferromagnetic phases of both $\rm RuO_2$~\cite{Berlijn2017} and $\rm MnF_2$~\cite{Erickson1953} belong to the spin point group $^24/^1m^2m^1m$. Following the symmetry analysis scheme in Ref.~\cite{Watanabe2024}, we derive the symmetry-allowed components of linear response tensors. Using the Kubo formula, the current $j^{\cal O}_i$ associated with an observable ${\cal O}$ in response to an external electric field $E_j$ is described by the ${\cal O}$-conductivity tensors $\sigma^{\cal O}_{ij}$, defined as
\begin{align}
    \sigma^{\cal O}_{ij} = \frac{e\hbar}{\pi V} \int dE~f(E)~{\rm ReTr} \left[ j^{\cal O}_i\frac{\partial G^R}{\partial E}v_j (G^R-G^A) \right], 
    \label{Kubo}
\end{align}
where $j^{\cal O}_i=(v_i{\cal O}+{\cal O}v_i)/2$ is the symmeterized ${\cal O}$-current operator, with the velocity operator $v_i=\partial\mathcal{H}/\partial(\hbar k_i)$, $V$ is the system volume, $f(E)$ is the Fermi-Dirac distribution function, and $G^{R/A}=1/(E-\mathcal{H}\pm i\Gamma)$ is the retarded/advanced Green function with the level broadening $\Gamma$. By substituting spin ($s_i$), electric quadrupole ($Q_{ij}$), and magnetic octupole ($M_{ijk}=s_iQ_{jk}$) operators into Eq.~(\ref{Kubo}), we identify nonvanishing components of the corresponding spin, electric quadrupole, and magnetic octupole conductivity tensors~\cite{SM}, as summarized in Table~\ref{TensorSpin}.

From this symmetry analysis (Table~\ref{TensorSpin}), we find that a rich variety of magnetic octupole and electric quadrupole currents emerges in $d$-wave altermagnets. First, we identify two distinct types of magnetic octupole Hall effects: (i) first-type magnetic octupole Hall effects (e.g., $\sigma^{M_{zu}}_{xy}$ and $\sigma^{M_{zv}}_{xy}$) that accompany the spin-splitter effect ($\sigma^{s_{z}}_{xy}$), and (ii) second-type magnetic octupole Hall effects (e.g., $\sigma^{M_{zyz}}_{xz}$ and $\sigma^{M_{zxz}}_{zy}$) that emerge even in the absence of the spin-splitter effect. The first type demonstrates that higher-order multipole Hall currents arise along with conventional spin-splitter currents, reflecting the orbital character embedded in spin-split wave functions. In contrast, the second type highlights that magnetic octupole degrees of freedom act as an independent transport channel, distinct from spin, and remain active even when spin-splitter effects are symmetry-forbidden. In addition, longitudinal magnetic octupole-polarized currents (e.g., $\sigma^{M_{zxy}}_{xx}$ and $\sigma^{M_{zxy}}_{zz}$) are generated, analogous to spin-polarized currents in ferromagnetic systems. These results collectively establish magnetic octupole moments as essential and active players in the nonequilibrium transport phenomena in $d$-wave altermagnets. 
Second, the $d$-wave electric quadrupole splitting [Fig.~\ref{fig:1}(d)] naturally gives rise to the electric quadrupole Hall effect, manifested as a nonzero $\sigma^{Q_{xy}}_{xy}$. This effect accompanies the spin-splitter effect ($\sigma^{s_{z}}_{xy}$) since spin and electric quadrupole splittings share the same symmetry [Fig.~\ref{fig:1}(c) and (d)].
This electric quadrupole Hall effect in $d$-wave altermagnets generates transverse currents that carry the same orbital quantity as in the orbital torsion Hall effect~\cite{Han2022} that is defined for centrosymmetric nonmagnetic systems.

In realistic materials, spin and orbital degrees of freedom are correlated through SOC. As a result, symmetry analysis must be performed with respect to the full magnetic point group of the system. Following the procedure outlined in Refs.~\cite{Seemann2015} and \cite{Zelezny2017}, we derive the ${\cal O}$-conductivity tensors in the relativistic limit (see Section III C in the Supplemental Material~\cite{SM}). Importantly, both types of magnetic octupole Hall effects remain symmetry-allowed even in the presence of SOC. In addition, we identify the symmetry properties of spin-orbital quadrupole conductivity tensors, where the spin-orbital quadrupole~\cite{Tang2023} is defined as the product of spin and orbital angular momentum, $\mathcal{Q}_{ij}=s_iL_j$. In contrast to conventional spin and orbital angular momentum currents, which preserve time-reversal symmetry, spin-orbital quadrupole currents inherently break time-reversal symmetry and reflect the underlying chiral electronic structure, thereby enabling unconventional magnetic orders~\cite{Mazzola2024,Fittipaldi2021}. The complete forms of the spin-orbital quadrupole conductivity tensors are provided in Table SVIII in the Supplemental Material~\cite{SM}.

\begin{figure}[t]
\begin{center}
\includegraphics[scale=0.48]{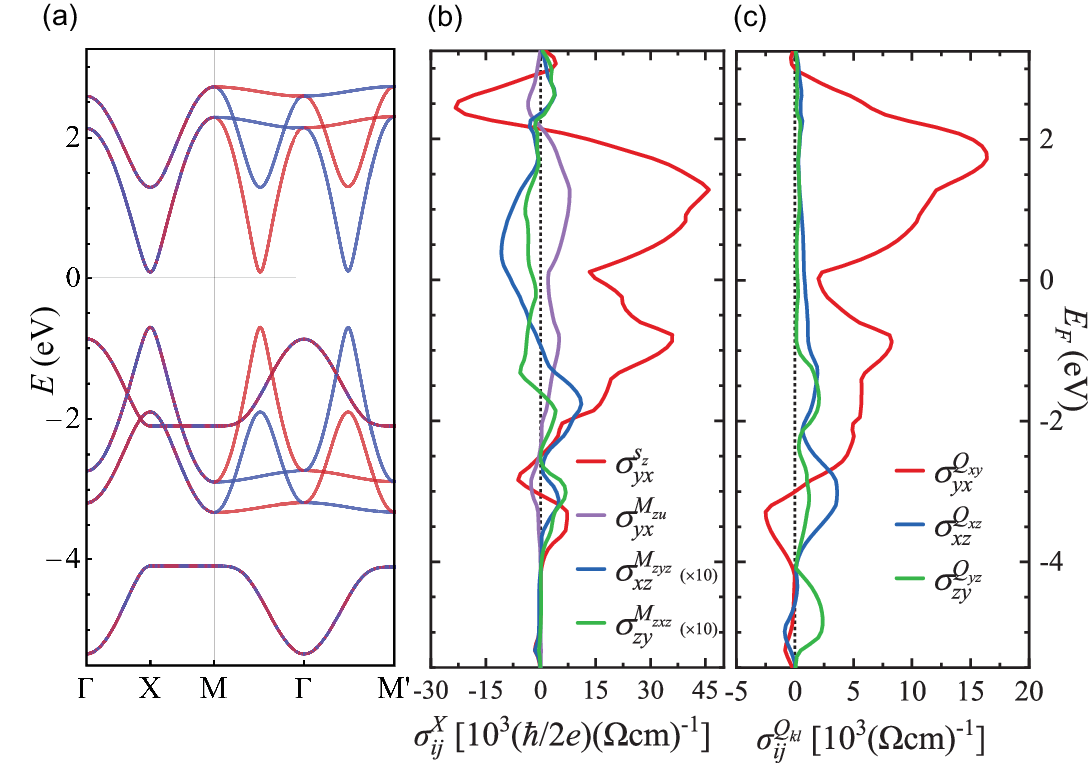}
\caption{
(a) Nonrelativistic band structure calculated from the Hamiltonian [Eq.~(\ref{Hbulk})], where majority and minority spin bands are shown in blue and red, respectively. (b) Spin and magnetic octupole Hall conductivities, and (c) electric quadrupole Hall conductivity, plotted as a function of Fermi energy $E_F$.
}
\label{fig:2}
\end{center}
\end{figure}

{\it Linear response calculation.}—We calculate the spin, electric quadrupole, and magnetic octupole conductivities in the nonrelativistic limit, based on the model Hamiltonian [Eq.~(\ref{Hbulk})], with a focus on the Fermi surface contributions~\cite{Bonbien2020,Freimuth2014,SM} that give rise to spin-splitter-like effects.
The calculated band structure [Fig.~\ref{fig:2}(a)] exhibits the characteristic $d$-wave spin splitting along the $\Gamma{\rm M}$ direction, consistent with the symmetry $k_xk_ys_z$~\cite{Bhowal2024}.

Several distinct features of the Hall responses in $d$-wave altermagnets are noteworthy. First, because spin and electric quadrupole splittings are intrinsically linked through the underlying orbital splitting [Figs.~\ref{fig:1}(c) and (d)], the spin-splitter effect ($\sigma^{s_z}_{yx}$) and the electric quadrupole Hall effect ($\sigma^{Q_{xy}}_{yx}$) always emerge simultaneously. For the same reason, $\sigma^{s_z}_{yx}$ and $\sigma^{Q_{xy}}_{yx}$ exhibit similar trends over a broad energy range [red curves in Figs.~\ref{fig:2}(b) and (c)].  

Second, the first-type magnetic octupole Hall effect ($\sigma_{yx}^{M_{zu}}$) emerges alongside the spin-splitter and electric quadrupole Hall effects. This Hall response originates from ferroic $Q_u$ ordering, induced by the tetragonal crystal symmetry~\cite{Bhowal2024}, in combination with the N\'eel order---magnetic octupole texture $\langle M_{zu}({\bf k})\rangle$ reproduces the pattern of spin splitting [Fig.~\ref{fig:1}(c)], as the homogeneous orbital character of $Q_u$ symmetry is imprinted on the wave functions. 
We note that this effect arises in altermagnets within the nonrelativisitic limit, in sharp contrast to a recent theory~\cite{Han2024} in which a magnetic octupole Hall effect
appears as a relativistic phenomenon driven by SOC in nonmagnetic systems. The nonrelativistic first-type magnetic octupole Hall effect identified in altermagnets is expected to be stronger in magnitude than the relativistic counterpart~\cite{Baek2025}.

\begin{figure}[b]
\begin{center}
\includegraphics[scale=1.0]{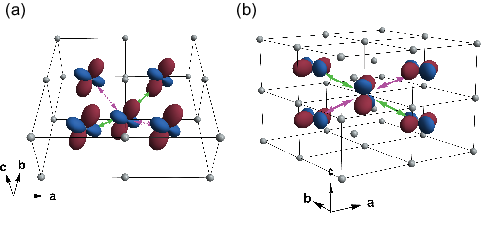}
\caption{
Schematic of intersublattice hopping between (a) $d_+(\equiv d_{yz}+d_{xz})$ orbitals and (b) $d_+$ and $d_{x^2-y^2}$ orbitals. Green and pink arrows indicate the sign of the hopping integrals, which depend on the relative phase of orbital wave functions along the hopping directions. In (a), the dashed arrows represent weaker hopping amplitudes compared to the solid arrows.
}
\label{fig:3}
\end{center}
\end{figure}

Third, the second-type magnetic octupole Hall effect ($\sigma^{M_{zyz}}_{xz}$ and $\sigma^{M_{zxz}}_{zy}$) emerges even when the spin-splitter effect is forbidden by symmetry. 
This Hall response originates from $d$-wave magnetic octupole textures in momentum space, induced by the interplay of hopping-mediated orbital hybridization with underlying spin and orbital orderings. To understand the origin, let us begin in the atomic limit, where all hopping is suppressed. In this limit, orbital and exchange splittings with identical spatial pattern, given by $\Delta\sigma_0\otimes\tau_z\otimes\{L_x,L_y\}+J\sigma_z\otimes\tau_z\otimes\mathbb{I}$, lead to atomic-level splittings that depend on spin, sublattice, and orbital character: 
\begin{align}
    \epsilon^\pm_{\sigma\tau}=\sigma\tau(J\pm\sigma\Delta),\quad
    \epsilon^{x^2\!-\!y^2}_{\sigma\tau}=\sigma\tau J,
\end{align}
where $\sigma\!=\!\pm$ denotes spin-$\uparrow$/$\downarrow$, $\tau\!=\!\pm$ labels the $M_1$/$M_2$ sublattice, and the superscript $\pm$ refers to the rotated orbitals $d_\pm\equiv d_{yz}\pm d_{xz}$ [Fig.~\ref{fig:1}(b)]. As an illustration, we consider two degenerate states, $|\!\uparrow\rangle\otimes|M_1\rangle\otimes|d_+\rangle$ and $|\!\downarrow\rangle\otimes|M_2\rangle\otimes|d_-\rangle$, which corresponds to $\epsilon^+_{++}=\epsilon^-_{--}$. 

When spin-independent intersublattice hopping is introduced (Fig.~\ref{fig:3}), the atomic orbitals hybridize in a momentum-dependent manner. These hopping processes fall into two categories: (i) \textit{intraorbital} hopping [Fig.~\ref{fig:3}(a)] and (ii) \textit{interorbital} hopping [Fig.~\ref{fig:3}(b)]. In the $ac$-plane, where the spin-splitter effect is symmetry-forbidden, the interorbital hoppings between $d_\pm$ and $d_{x^2\!-\!y^2}$ orbitals acquire a $d$-wave form factor $\pm k_xk_z$, while intraorbital hoppings remain isotropic. Consequently, the anisotropic hybridization of the two degenerate atomic states with $d_{x^2\!-\!y^2}$ orbital generates a nonvanishing $d$-wave electric quadrupole texture $\langle Q_{xz}({\bf k})\rangle$ [Fig.~S1(a)~\cite{SM}], reflecting the superposition of $d_{xz}$ and $d_{x^2\!-\!y^2}$ orbitals, with $\langle Q_{yz}({\bf k})\rangle$ being compensated. Due to the opposite spin character of the two hybridized states, concomitant magnetic octupole textures also emerge: a compensated $\langle M_{zxz}({\bf k})\rangle$ and a nonvanishing $\langle M_{zyz}({\bf k})\rangle$ [Fig.~S1(b)~\cite{SM}], both sharing the $d$-wave symmetry. The nonvanishing magnetic octupole texture indicates spin-dependent hybridization between $d_{yz}$ and $d_{x^2\!-\!y^2}$ orbitals. In the absence of orbital splitting (i.e., $\Delta=0$), the fourfold degeneracy $\epsilon^+_{++}=\epsilon^-_{++}=\epsilon^+_{--}=\epsilon^-_{--}$ prevents the formation of any net magnetic octupole texture, although the electric quadrupole texture still persists (see Section V in the Supplemental Material~\cite{SM}). 

When an electric field is applied along the nodal direction of the magnetic octupole textures, a finite magnetic octupole Hall response arises—analogous in form to the spin-splitter effect, yet fundamentally distinct in origin, as it does not rely on spin-split band structure. This mechanism highlights a novel type of transport driven by multipolar orbital degrees of freedom, rather than conventional spin-based mechanisms.
A similar analysis applies to the second-type magnetic octupole Hall components in the $bc$-plane. For a detailed discussion, refer to the Supplemental Material~\cite{SM}.

{\it Discussion and outlook.}—We have demonstrated the multipole Hall effect in $d$-wave altermagnets, which host ferroic magnetic octupole order. The magnetic octupole Hall effect represents a dintinct and indispensable addition to the family of Hall phenomena in altermagnets, extending the landscape of Hall effects to include higher-rank multipole transport~\cite{Tahir2023}. Importantly, the unique symmetry characteristics of magnetic octupole moments enable a symmetry-based distinction from conventional spin-based Hall effects, providing an unambiguous root for experimental identification.

Our findings suggest several promising experimental schemes for probing multipolar degrees of freedom in altermagnets, both in equilibrium and nonequilibrium regimes. In equilibrium, magnetic multipolar order can be accessed via spin-resolved and polarization-dependent angle-resolved photoemission spectroscopy~\cite{Zhang2013,Xie2014}. The exact momentum-space coincidence of spin- and orbital-split electronic bands would serve as a spectroscopic signature of the simultaneous formation of spin and orbital orderings, signaling the spontaneous emergence of magnetic multipole moments. 

In nonequilibrium, the generation of multipole currents opens new avenues for current-induced magnetization control beyond traditional spin-torque mechanisms~\cite{Bai2022,Karube2022}. Just as multipole exchange interactions stabilize local multipolar orders in equilibrium~\cite{Kusunose2008,Santini2009,Pi2014,Pourovskii2016,Iwahara2022,Mosca2022}, an exchange coupling between itinerant and localized multipole moments is symmetry-allowed. This coupling enables the transfer of multipole moments between them—a higher-order analogue of conventional spin-transfer torque~\cite{Slonczewski1996,Berger1996}. One promising experimental approach is through current-induced magnetic torque measurements in altermagnetic spin valves, where multipolar currents injected from one altermagnet exert torque on an adjacent altermagnet. 

Complementary to octupolar torque mechanisms, orbital dynamics provide another pathway for magnetization control. The sizable electric quadrupole Hall effect observed in $d$-wave altermagnets highlights the central role of orbital degrees of freedom. Recent theoretical work~\cite{Han2025} has shown that currents of orbital angular position—equivalently, electric quadrupole currents—can be pumped by magnetization dynamics. In the reciprocal process, the injection of electric quadrupole currents can exert torques on the magnetic order. Thus, the pronounced electric quadrupole Hall effect in $d$-wave altermagnets [Fig.~\ref{fig:2}(c)] offers an efficient mechanism for current-induced control via orbital channels. 

The significance of multipolar order in altermagnets positions them as promising candidates for interdisciplinary research, spanning multiferroics~\cite{Fiebig2016}, strongly correlated electron systems~\cite{Kusunose2008,Kuramoto2009,Santini2009}, and emerging fields such as multipolectronics~\cite{Tahir2023}. Recently, a new class of multiferroic materials has been proposed, combining (anti)ferroelectricity with altermagnetism~\cite{Duan2025,Gu2025,Smejkal2024}. These systems enable electric-field control of altermagnetic order through switching of electric polarization. In Mott insulators with quadrupolar and octupolar orderings, electric polarization can be induced by the higher-order multipoles~\cite{Banerjee2025,Zhao2025}, leading to a concept of \textit{multipolar multiferroicity}---an extension beyond conventional dipolar cases---which may also apply to altermagnets. Additionally, resonant phonon modes have been shown to drive multipolar dynamics in the Mott regime and even reverse the sign of octupole moment~\cite{Hart2025}. This broadens the scope of magnetization dynamics in altermagnets from dipolar~\cite{Gomonay2024} to multipolar framework.

Finally, we note that the relevance of multipole order is not limited to $d$-wave altermagnets, but extends to $g$- and $i$-wave altermagnets, where even richer symmetry-allowed multipole configurations can emerge~\cite{McClarty2024,Verbeek2024,Belashchenko2025,GalindezRuales2023}. In addition to identifying unconventional symmetry-breaking phases and emergent quantum phenomena, altermagnets provide a fertile ground for exploring the nonequilibrium dynamics of multipole moments, opening new frontiers in condensed matter physics.

{\it Acknowledgments.}---We thank Changyoung Kim and Yeong Kwan Kim for fruitful discussions. This work was supported by the National Research Foundation of Korea (NRF) grant funded by the Korea government (RS-2024-00436660).

\end{document}